\documentclass[preprintnumbers,amsmath,amssymb]{revtex4}
\begin{document}
\title{Comment on ``Dipole transitions and Stark effect in the charge-dyon system,''\\ by L. Mardoyan et al. \cite{Mardoyan:2006qd}}
\author{E. A. Tolkachev and L. M. Tomilchik}
\altaffiliation{email address: tolkachev, tomilchik@dragon.bas-net.by}
\affiliation{Institute of Physics, 68 Nezavisimosti Avenue, Minsk 220072, Belarus}
\date{Oct. 4, 2006}
\begin{abstract}
\end{abstract}
\maketitle
In a recent e-print ({\tt cond-mat/0609768}, ``Dipole transitions and Stark effect in the charge-dyon system'', by L.~Mardoyan, A.~Nersessian, H.~Sarkisyan, V.~Yeghikyan, \cite{Mardoyan:2006qd}) the authors claim that "{\it the presence of monopole, besides providing the system with degenerated (on azimuth quantum number) ground state, makes possible the dipole transitions         which obey selection rules $(l'=l; m'=m, m'=m\pm 1)$, where $l,m$ are          respectively, orbital and azimuth quantum numbers}".
Also they calculated the linear and quadratic Stark effect in such a system and pointed out that "{\it the linear Stark effect in the ground state is proportional to azimuth quantum number, and to the sign of monopole number}" and "{\it the quadratic Stark effect in the ground state is independent on the signs of azimuth and monopole numbers}".

It is well known \cite{Tolkachev:1983st,Tolkachev:1988mn}
that modification of the selection rules in a
theory with monopole arises as a straightforward consequence of the
non-invariance of model under spatial reflection (P) which is a by-product of consistent consideration of magnetic monopoles, both Abelian and non-Abelian
\cite{Tomilchik:1963,Strazhev:1975,Goddard:1977da}.

It was also demonstrated on a simple model example of a bound non-relativistic system electron-dyon, which is described by Schr\"odinger or Pauli Hamiltonian, that the selection rules for a dipole radiation, unlike the familiar  quantum-mechanical situation, admit the transitions with $\Delta l =0$ \cite{Tolkachev:1988mn}. It was pointed out that in the consistent relativistic theory based on Dirac equation the same modification is taking place as well \cite{Kazama:1976fm,Kazama:1977um,Zhang:1990fk,Zhang:1988ab}.

Consideration of properties of the bound system charge-monopole or charge-dyon in external fields also allows to observe the above mentioned effects, which are   related with P-parity violation. In particular, Stark and Zeeman effects, as well as photoionization of the ground state of the  bound system charge-dyon was analyzed in \cite{Shnir:1992iz,Tolkachev:1989cr,Tolkachev:1990rw}. Since the modification of the selection rules for a dipole radiation arises from angular part of the Hamiltonian and is related with  matrix element sandwiched between generalized angular harmonics,  inclusion of additional radial-dependent interaction terms, does not affect this result. It particular, the spectrum of Pauli and Dirac Hamiltonian of such a system with usual Coulomb electromagnetic potential and corresponding matrix elements of the dipole moment operator were considered by \cite{Bose:1985yr,Bose:1986mc,Zhang:2002rk}.

The inclusion of the extra radial dependent term $s^2/2 r^2$ into the non-relativistic Hamiltonian of the charge-dyon system does not make any principal difference comparing to the previous consideration. At least, the authors have to refer the related works properly.

\end{document}